\pdfoutput=1

\title{Predictive coding: A Possible Explanation of Filling-in at the blind spot }
\author{Rajani Raman, Sandip Sarkar\\ \footnotesize{Saha Institute of Nuclear Phycsis, Kolkata, India}}

\date{}

\documentclass[12pt]{article}
\usepackage{amsmath,amssymb}
\usepackage{cite}
\usepackage{graphicx}
\usepackage{nameref,hyperref}
\usepackage[font={small}]{caption}

\begin{document}
\maketitle

\begin{abstract}
Filling-in at the blind-spot is a perceptual phenomenon in which the visual system fills the informational void, which arises due to the absence of retinal input corresponding to the optic disc, with surrounding visual attributes. Though there are enough evidence to conclude that some kind of neural computation is involved in filling-in at the blind spot especially in the early visual cortex, the knowledge of the actual computational mechanism is far from complete. We have investigated the bar experiments and the associated filling-in phenomenon in the light of the hierarchical predictive coding framework, where the blind-spot was represented by the absence of early feed-forward connection. We recorded the responses of predictive estimator neurons at the blind-spot region in the V1 area of our three level (LGN-V1-V2) model network. These responses are in agreement with the results of earlier physiological studies and using the generative model we also showed that these response profiles indeed represent the filling-in completion. These demonstrate that predictive coding framework could account for the filling-in phenomena observed in several psychophysical and physiological experiments involving bar stimuli. These results suggest that the filling-in could naturally arise from the computational principle of hierarchical predictive coding (HPC) of natural images.
\end{abstract}

\section*{Introduction}

Filling-in at the blind spot is one of the examples of how brain interpolates the informational void due to a deficit of visual input from the retina. Due to the absence of photoreceptors at optics disc, the retina is unable to send the corresponding signal to the brain and thereby, hides some portion of the visual field. This concealed visual field is known as the blind spot. However, we never notice any odd patch in our visual field, even in monocular vision, but rather we see the complete scene; filled up in accordance with the surrounding visual attributes~\cite{ramachandran1992blind}. This completion is known as perceptual filling-in or simply filling-in. In addition to the blind spot, filling-in also occurs in other visual input deficit conditions,  like filling-in at the artificial and natural retinal scotoma~\cite{ramachandran1991perceptual,gerrits1969filling}.  Besides the deficit of input, filling-in also occurs in visual illusions such as Neon color spreading, Craik-O'Brien-Cornsweet illusion, Kanizsa shapes, etc.  and steady fixation condition like Troxler effect (for review see~\cite{komatsu2006neural}).

\begin{figure}[]
\includegraphics[width=5.3in]{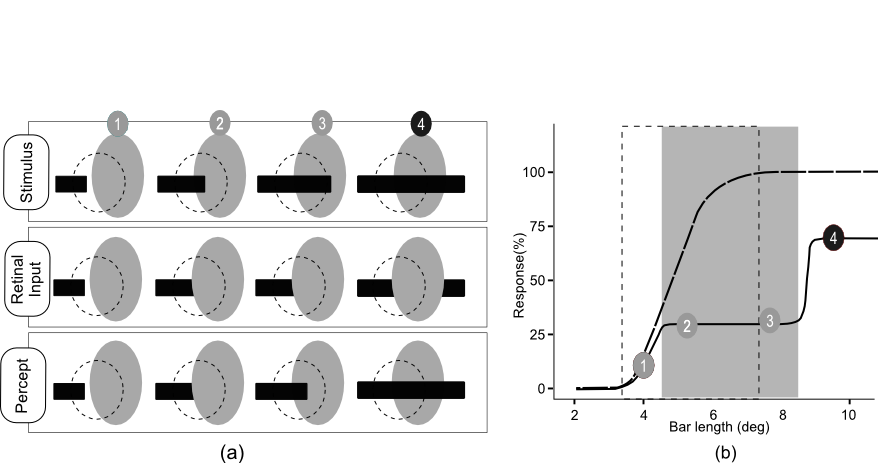}
\caption{{\bf Schematic Illustration of bar completion experiment}(adopted from Matsumoto and Komatsu ~\cite{komatsu2006neural}).
(a) The gray oval area represents the blind spot, whereas the dashed circle represents the receptive field of a neuron. The actual stimulus and corresponding retinal input and percept at position 1,2,3 and 4 has shown. Stimuli, here, is the bar with one end fixed outside and the other end is drifting across the blind spot. (b) The response of a typical neuron in BS region at the deep layer of primary visual cortex. The gray rectangle indicates the blind spot and the dotted rectangular area represent the receptive field of the typical neuron. The solid line is the response obtained through the eye that provides the blind spot (BS eye) and the dotted line is the response of the same neuron obtained through the fellow eye. While the drifting end of the bar was inside the blind spot the perception of the bar was of a short isolated bar and corresponding neural responses were low and constant. But the moment bar end crossed the blind spot, the neural response elevated rapidly and completion of the bar was perceived. These elevated response exhibit nonlinearity; the response to the long bar that stimulate simultaneously the both sides of the blind spot was larger than the sum of responses to the stimuli either side of blind spot separately.}
\label{Komatsu_Bar_Experiment}
\end{figure}

Many psychophysical and physiological studies have been performed to gain insight into the neural mechanism of perceptual completion at the blind spot ~\cite{junior1992dynamic,komatsu2000neural,matsumoto2005neural}. The two main findings of these studies have been: first, filling-in is an active phenomena, where the visual system, rather than being idle in the absence of visual information, is involved in some neural computation, and second, activities in early visual cortical areas are involved in filling-in process. Physiological studies on the monkeys show that  perceptually correlated neural activities are evoked in  the deep layer of primary visual cortex, in the region that retinotopically corresponds to the blind spot (BS region), when filling-in completion occurs~\cite{komatsu2000neural,junior1992dynamic}. In an experiment, Matsumoto and Komatsu~\cite{matsumoto2005neural} showed that some active neuron in BS region in deep layer of primary visual cortex (BS neurons), which possess larger receptive fields that extend beyond the blind spot, exhibits non-linear elevated response when a long moving bar cross over the blind spot and perceptual completion occurs~(See fig.\ref{Komatsu_Bar_Experiment}).

Although some initial attempts have been made to understand the computational mechanism of completion of illusory contour and surface ~\cite{neumann2001visual,neumann1999recurrent,grossberg1985neural,grossberg1999neural,grossberg1997visual}, little work has been devoted to the study of computational mechanism of filling-in completion at the blind spot. Recent studies ~\cite{satoh2008computational} have suggested the computational mechanism of completion of the bar  in terms of a complex interaction of velocity-dependent pathways in the visual cortex under the framework of regularization theory. But the fitness of his proposal in the context of a general coding principle of the visual cortex is not clear. Here in this study, we suggest the filling-in completion at the blind spot naturally follows from the mechanism of Hierarchical predictive coding (HPC)of natural images, which has been, recently, gained growing support as general coding principle of visual cortex ~\cite{rao1999predictive,lee2003hierarchical,friston2005theory,hohwy2008predictive,jehee2009predictive,jehee2006learning,huang2011predictive,clark2013whatever}.

The root of Hierarchical predictive coding lies in the probabilistic hierarchical generative model and the efficient coding of natural images. In such probabilistic frameworks, the job of the visual system is to infer or estimate the properties of the world from signals coming from receptors ~\cite{mumford1994neuronal,dayan1995helmholtz,olshausen1996emergence}. In HPC framework, this job is hypothesized to be completed by concurrent prediction-correction mechanism along the hierarchy of the visual system. Accordingly, each higher visual area (say V2) attempt to predict response at its lower area (say V1) on the basis of the learned statistical regularities, and send that prediction signal to the lower area by feedback connection. In response to this top-down information, lower area sends a residual error signal to the higher area, by feed-forward connection, to correct the next prediction. This idea is based on the anatomical architecture of the visual system that is hierarchically organized and reciprocally connected ~\cite{felleman1991distributed}. Probabilistic generative model, in HPC framework,  accounts for learning the statistical regularities found in natural images and generation of prediction of input  based on that learning. Recently, several neuronal tuning properties in different visual area such as the lateral geniculate nucleus (LGN), primary visual cortex (V1) and  middle temporal level (MT) has been explained using this framework ~\cite{jehee2006learning,jehee2009predictive}. For example, Rao~\cite{rao1999predictive} suggested that the extra-classical properties of neurons in V1 could be understood in terms of predictive-feedback signal from the secondary visual cortex (V2) that is made in the larger context and the backdrop of learned statistical regularity from natural Scene. We speculated that the similar mechanism could also explain the filling-in completion across the blind spot.

In this work, we have conducted simulation studies involving horizontal bars on three leveled (LGN-V1-V2) HPC model network having a blind spot which was emulated by removing the feed-forward (LGN-V1) connection. In our first investigation we have employed shifting bar stimuli as described in~\cite{matsumoto2005neural}(See fig. \ref{Komatsu_Bar_Experiment}), to study the properties of our model network and recorded the model predictive estimator neurons (PE neurons) in V1 in BS region. We found that these neurons exhibit similar non-linear response and represents the filling-in completion when bar crosses the blind spot. In another investigation, we presented two separate bar segment at the opposite end of the model blind spot to verify the tolerance of completion by varying the alignment of those segments. We found that the filling-in completion is best when the bars are perfectly aligned. The completion is visible for small orders of misalignment, but it fades out quickly with increasing misalign. These results are  consistent with the finding of psychophysical experiments~\cite{araragi2011anisotropies,araragi2008anisotropy}. These findings suggest that the filling-in process could naturally arise in the computational principle of hierarchical predictive coding (HPC) of natural images which has been, recently, argued to be a general coding principle of the visual system.

\section*{Model}

\subsection*{Hierarchical Predictive coding of natural images:}

\begin{figure}
\includegraphics[width=5.3in]{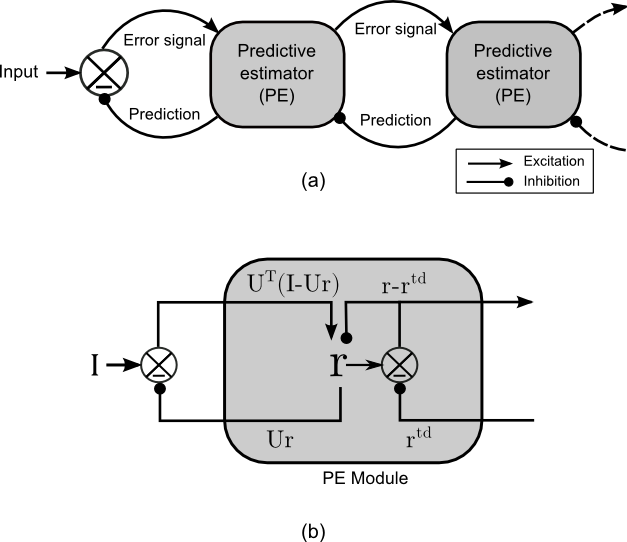}
\caption{{\bf General  HPC Architecture.}
a) On arrival of input, predictive estimator module at each higher visual processing level makes the estimate and sends prediction signal to its next lower level by feedback connection and receives the corresponding prediction error by a feed-forward connection. The error signal is used by the predictive estimator to correct the estimate for better prediction. b) General predictive estimator (PE) module constitutes of (i) neurons to represent the estimate of the input I by their response vector r by minimizing the bottom-up ($\mathbf{I}-U \mathbf r$) and top-down ($\mathbf{r} - \mathbf{r}^{td}$) error, (ii) feed-forward error carrying neurons has the efficacy matrix $U$, which encode the basis vectors their synaptic weights (or receptive fields), (iii) prediction $U \mathbf r$ carrying neurons and (iv) top down error detecting neutrons. }
\label{Genaral_HPC}
\end{figure}

\paragraph*{ General Model and Network Architecture:} As indicated in the previous section, the problem of vision, in a probabilistic framework, has been considered as an inference or an estimation problem; where an organism try to estimate the hidden physical cause (object attributes such as shape, texture and luminance etc.) behind the generated image that organism receives as an input. In line with these suggestions, it is assumed that, in HPC framework, image generation involves the interaction between physical causes at multiple spatial and temporal level along the hierarchy. The goal of the visual system is, thus, to estimate (or internally represent) these hidden causes at different levels of visual processing, efficiently ~\cite{rao1999predictive}. This goal is, in HPC framework, hypothesized to be achieved by the visual system using recurrent predictions-corrections mechanism throughout the hierarchy ~ (See fig. \ref{Genaral_HPC}a).

In this framework, on the arrival of an input, predictor estimator modules (PE module) at each visual processing level generate the prediction (or estimate) on the basis of the learned statistical regularities of natural scenes. Each higher area (say V2) then sends these generated prediction to its immediate lower level (say V1) by feedback connections and in return receives the error signal, by feed-forward connections, which is used to correct the current estimate. An equilibrium state is achieved after concurrent prediction-correction cycle; where the estimate matches the input signal. This optimum-estimate is regarded as a representation of the input at that level. The optimum-estimate achieved at different levels of network is depicted as a perception corresponding to that image. 

In general, a single PE module ~ (See fig. \ref{Genaral_HPC}b) consist of: (i) Predictive estimator neurons (PE neurons) which represent the estimate of current input signal $\mathbf I$ with response vector $ \mathbf r $ (state vector), (ii) neurons, carrying prediction signal $U \mathbf r$ (for the input $\mathbf I $) to lower level by feed-back connections, whose synapse encode encoding efficacy matrix $U$, (iii) neurons, carrying feed-forward error signal ($\mathbf{I}-U \mathbf r$) form level 0 to level 1, whose synapses encoded rows of efficacy matrix $U^T$,  and (iv) error detecting neurons which carry the  residual error signal $\mathbf{r} - \mathbf{r}^{td}$ to the higher level.

\paragraph* {Network dynamics and learning rule:} The dynamics, the learning rules and hence the above-mentioned architecture of a general PE module directly stem from probabilistic estimation methods. In the Bayesian framework, these originate from maximum a posteriori (MAP) approach. In this case, maximizing the posterior probability $P(\mathbf r,\mathbf{r}^{td},U| \mathbf I) $, which is equal to the product of generative models $ P(\mathbf {I}|\mathbf{r},U)$, $P(\mathbf{r}^{td}|\mathbf{r})$ and prior probabilities $P(\mathbf{r})$ and $P(U)$, with respect to $\mathbf{r}$ and $U$ provides the dynamics and learning rule respectively. Equivalently, in the framework of information theory, the  minimum description length (MDL) approach leads to same results by minimizing the coding length $E$, which is equal to negative log of posterior probability defined above, with respect to $\mathbf{r}$ and $U$ (for details, see \cite{ballard2012dynamic,rao1999predictive}).

By assuming the probability distributions $P(\mathbf {I}|\mathbf{r},U)$ and $P(\mathbf{r}^{td}|\mathbf{r})$ as Gaussians of zero mean and variances $\sigma^2$ and $\sigma_{td}^2$ respectively, the  total coding length $E$ can be written as,

\begin{equation}\label{eq:cost}
    E=\dfrac{1}{\sigma^2}(\mathbf{I}-U\mathbf{r})^T(\mathbf{I}-U\mathbf{r})+\dfrac{1}{\sigma_{td}^2}(\mathbf{r}-\mathbf{r}^{td})^T(\mathbf{r}-\mathbf{r}^{td})+g(\mathbf r)+h(U)
\end{equation}

here, $g(\mathbf r)$ and $h(U)$ are the negative log of prior probabilities $P(\mathbf r)$ and $P(U)$ respectively. Minimizing the coding length $E$, with respect to $\mathbf r$ (using the gradient descent method) provides the dynamics of PE module as,

\begin{equation}\label{eq:dynamics}
    \frac{d\mathbf{r}}{dt}=-\dfrac{k_1}{2} \frac{\partial E} {\partial \mathbf r}=\frac{k_1}{\sigma^2}U^T (\mathbf{I}-U \mathbf{r}) + \frac{k_1}{\sigma_{td}^2}( \mathbf{r}^{td}- \mathbf{r})- \frac{k_1}{2} g'(\mathbf r)
\end{equation}

here, $k_1 $ is a rate parameter that governs the rate of descent towards a minimum of $E$, and $U^T$ is the transpose to weight matrix $U$. The steady state of this dynamical equation provides an optimum-estimate, which is regarded as the representation of the input. Coding length $E$, roughly, can be seen as the mean square error at the input and the output level of a PE module, subjected to constraints of prior probabilities. And minimization of the coding length is equivalent to optimization of estimate by recurrently matching of estimate to the corresponding ``sensory driven'' input from lower area as well as ``context driven'' prediction signal from higher area. The prediction signal $U \mathbf r$  is the linear combination of basis vectors $U_i$`s. The $U_i$ is the $i^th$ column of the matrix $U$, and represents the receptive field for $i^{th}$ neuron. The weighted coefficient in this combination, $r_i$, represents the response of $i^{th}$ neuron having receptive field $U_i$. The visual representation of the predication $U\mathbf r$ corresponding to optimum-estimate $\mathbf r$ is, in this study, termed as "perceptual image."

Furthermore, the minimization of coding length $E$, with respect to $U$ using gradient descent method provides the learning rule for basis matrix $U$ as,

\begin{equation}\label{learn}
    \frac{dU}{dt}=-\dfrac{k_2}{2}\frac{\partial E}{ \partial U}=\frac{k_2}{\sigma^2} ( \mathbf{I}-U\mathbf{r})\mathbf{r^T}- \frac{k_2}{2} h'(U)
\end{equation}

here $k_2$ is learning rate, which operates on the slower time scale than the rate parameters $k_1 $, and $\mathbf r^T$  is the transpose of state vector $\mathbf r$. This learning rule can be seen as of Hebbian type. In this study, prior probability, $P(\mathbf r)$, on state vector $\mathbf r$, are chosen according to sparse coding; where it is assumed that the visual system encodes any incoming signal with a small set of neurons from the available larger pool. The kurtosis prior distribution ($P(r_i)=exp(-\alpha log(1+r_i^2))$) constrains the dynamics for the sparse representation of the input. This distribution gives us:

\begin{equation}\label{eq:gdash}
    g'(r_i)=2 \alpha r_i / (1+r_i^2)
\end{equation}

which is used in equation~(\ref{eq:dynamics}). The prior probability distribution, $P(U)$ has been chosen here to be Gaussian type, which finally gives us:

\begin{equation}\label{eq:gdash}
h'(U)= 2 \lambda U 
\end{equation}

 which is used in the equation ~(\ref{learn}). Here $\alpha $  and $\lambda $ are variance related parameters.

\subsection*{Simulation}

\begin{figure}
\includegraphics[width=5.3in]{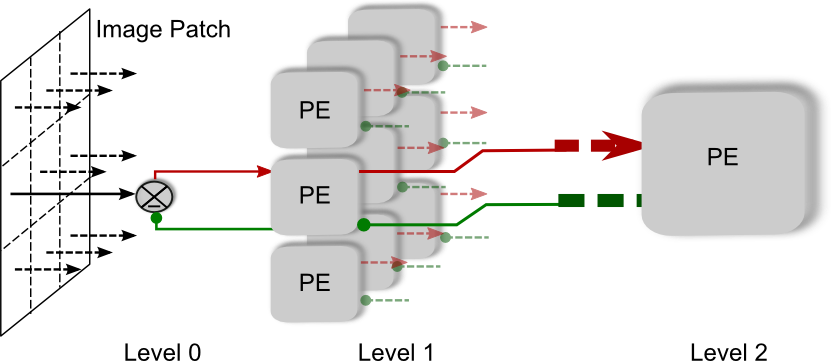}
\caption{{\bf Three level HPC model network.}
 Each of nine level 1 PE module sends prediction to level 0 by feedback connection and receives error signal corresponding to their own local image patches by a feed-forward connection. On the other hand, the sole PE module at level 2 sends the prediction signal to all level 1 modules and in reply, receives the error signal collectively from all these modules. Level 2, therefore, encodes larger visual patch and hence possess the larger receptive field. }
\label{3_level_HPC}
\end{figure}

\paragraph*{Network } In this work we simulated a three level linear hierarchical predictive network~(See fig.\ref{3_level_HPC}). In this network, Level 1, which is equivalent to V1,  consists of 9 predictor estimator modules (PE modules). These modules receive input from level 0 and send the output to the sole module at level 2. Level 0 is equivalent to the LGN and level 2 is equivalent to V2. Therefore, the PE module at level 2 receives input from all the nine level 1 PE modules and sends back the feedback signal to all of them. This architecture is based on the fact that the visual area higher in hierarchy operates on a higher spatial scale.

Each of nine PE modules at level 1 consists of 64 PE neurons, 144 Prediction carrying neuron, 64 afferent error carrying neurons and 64 error detecting neurons for conveying the residual error to level 2. The layer 2 module consists of 156  PE neurons, 576 prediction carrying neurons and 156 error carrying neurons.


\begin{figure}
\includegraphics[width=5.3in]{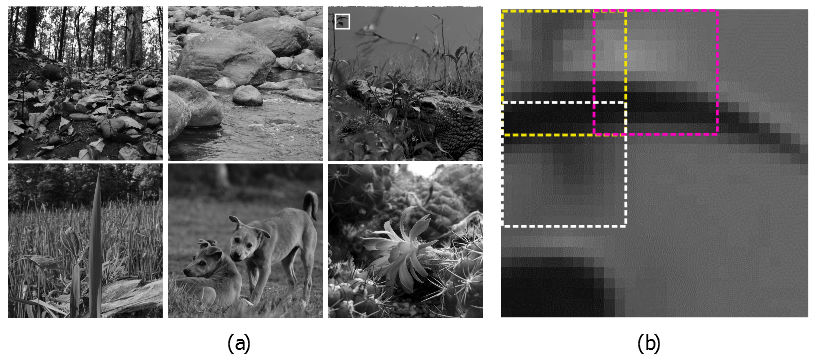}
\caption{{\bf Natural Images.}
a) These images, taken from of different natural environments, are used for simulation. b) A typical sample of 30 x 30 pixels image patches extracted from the natural image (top rightmost) from the position shown by the white rectangle. Each of these patches is broken down to 9 sub-patches of $12 \times 12 $ pixel each with 3 overlapping pixels. Three such sub-patches are shown here by three dotted rectangles in yellow, magenta and white.
Each of these sub-patches forms the local input to the 9, level 0 modules in the HPC model network. 
}
\label{Natural_Images}
\end{figure}

\paragraph*{Training:}  Six natural images (fig \ref{Natural_Images}a) of size $512 \times 512$ pixels were used for training after pre-processing. The pre-processing involved  DC removal and the filtering of images with circular symmetric  whitening/lowpass filter with spatial frequency profile $W(f)= f exp(-(f/f_0)^4)$( see ~\cite{olshausen1996emergence}). Cutoff frequency $f_0$ was taken to be 200 cycles/image. The pre-processing has been argued to emulate the filtering at LGN~\cite{atick1992could}. Variance normalized 1000 batches of 100 image patches of size $30 \times 30$ pixel, which were extracted from randomly selected locations from the randomly selected pre-processed images, were given as input to the network. A single $30 \times 30$-pixel image consisted of nine tiled $12 \times 12$-pixel image patches, which were overlapped by 3 pixels (see fig.\ref{Natural_Images}b) and which were fed to the corresponding level 1 PE modules. For each batch of image patches, the network was allowed to achieve steady states (according to the equ. ~(\ref{eq:dynamics})) and the average of these states was used to update the efficacy of neurons~( according to the equ. \ref{learn}), initially assigned to random values. During training the gain of efficacy vectors (rows of $U^T$ or columns of $U$) was adopted so as to maintain equal variance on $r_i$. The level 1 was trained first and then the level 2. Parameter values used in this study are: $k_1= 1$, $k_2= 3$, $\sigma^2= 3 $, $\sigma_{td}^2 = 10 $, $\alpha= 0.05$ at level 1 and $ 0.1 $ at level 2, $\lambda= 0.0025$

To simulate the deficit of input related to optic disc (or blind spot), the efficacy of   early feed-forward (level 0 - level 1) neurons, that carry the error signal corresponding to the middle region (of size $ 8 \times 8 $) of input patches (of  size $ 30 \times 30 $), were set to zero. 


\section*{Results}

\begin{figure}
\includegraphics[width=5.3in]{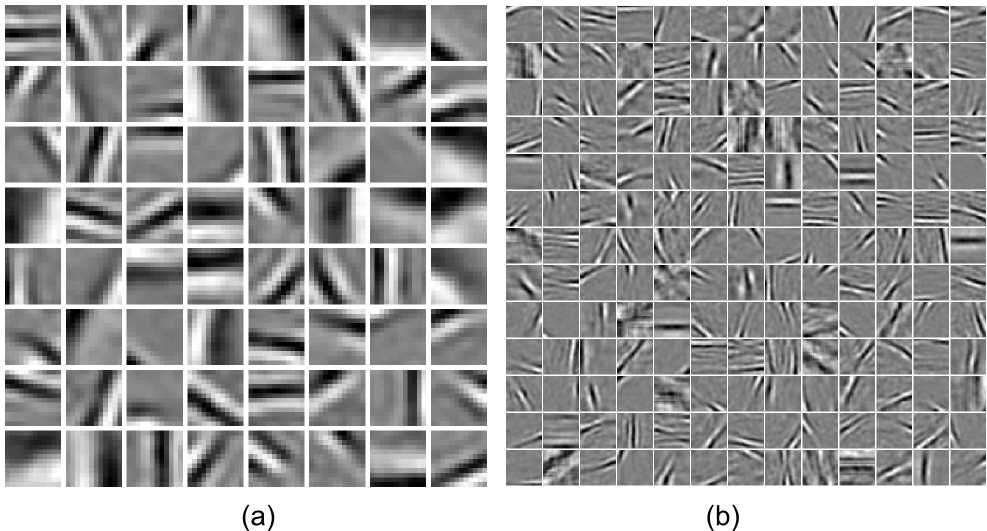}
\caption{{\bf Learned synaptic weight:} Learned synaptic weights of feed-forward model neurons, achieved with sparse prior distribution on network activities after training with natural image patches,  at level 1 and level 2 are presented in (a) and (b) respectively.}
\label{Receptive_field}
\end{figure}

To ascertain whether the computational mechanism of HPC could account for filling-in completion across the blind spot, we conducted a pair of experiments using bar stimuli on the trained HPC model network. The HPC network was allowed to learn the synaptic weight of model neurons, by exposing it to natural image patches under the constraints of the sparseness of model neuron responses (see method). The learned synaptic weights of neurons carrying feed-forward signal of one of the modules at level 1 and level 2 are shown in fig.~\ref{Receptive_field}. The weighting profiles at level 1~(fig.\ref{Receptive_field}a) resemble the Gabor-like receptive field at V1, which is similar to the results reported earlier in several studies\cite{rao1999predictive,jehee2009predictive,olshausen1996emergence}. The weighting profile at the level 2~(fig.\ref{Receptive_field}b) resembles the more abstract visual feature like the long bar, curve, etc..The blind spot was emulated in the network by removing feed-forward connection (see method), whereas, the training was performed on a network with feed-forward connection. We designate the network with the blind spot as BS network and the one without the blind spot as a non-BS network. 

\begin{figure}
\begin{center}
\includegraphics[width=5.3in]{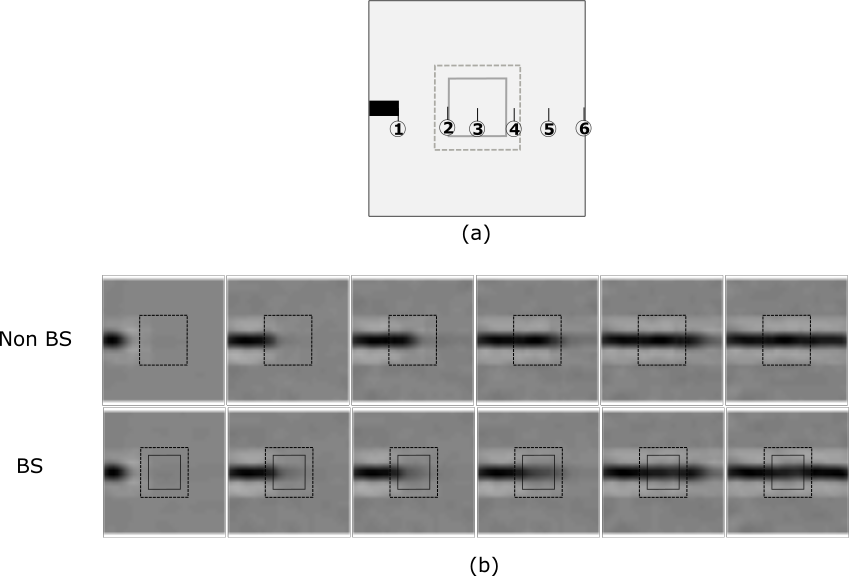}
\end{center}
\caption{{\bf Bar Shift Exp.} a) A typical $30 \times 30 $ pixel stimulus is shown here. The darkened object in the stimulus is a bar, whose endpoint is represented by the number 1. Five more stimuli were constructed by shifting the bar end to positions 2 to 6. The larger rectangle of size $12 \times 12 $ pixels (shown by the dotted line at the center) indicates the extension of BS module and the smaller one of size $8 \times 8$ (shown by the solid line) indicate the extension of blind-spot. b) Generated $30 \times 30 $ ``perceptual images'' corresponding to response profile of PE neurons at level 1 of the HPC network for non-BS (top row) and BS (bottom row) cases are shown. }
\label{Bar_shift}
\end{figure}

\paragraph*{ Filling-in of shifting bar :} Both BS and non-BS Network were exposed to images of a horizontal bar of different length. One end of the bar was fixed at a position outside of the blind spot, whereas, the position of other end was varied (by one pixel each instant) across the blind spot. Pictures of the bar for six different end positions are shown in figure (\ref{Bar_shift}a). The response vector, $\mathbf{r}$, of PE neurons in the central module in the model network (let say BS module) at level 1 was recorded for the different end position of the bar.

\begin{figure}
\begin{center}
\includegraphics[width=5.3in]{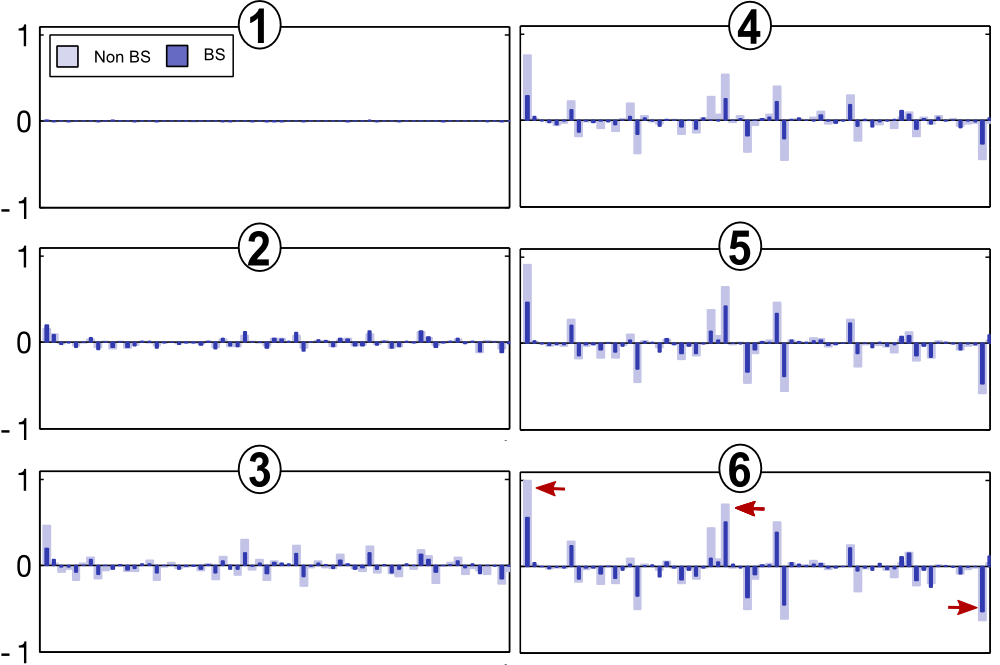}
\end{center}
\caption{{\bf Responses profiles:}
Normalized responses of 64 PE neurons at BS module, corresponding six stimuli discussed in figure 6(a) are presented. The dark blue bar represents the response of PE neurons for the BS network, whereas, the light blue bar represents the responses for the non-BS network. Three most highly active neurons (in bottom leftmost bar plot) are marked by red arrows. }
\label{Response_Barshift}
\end{figure}

\begin{figure}
\begin{center}
\includegraphics[width=5.3in]{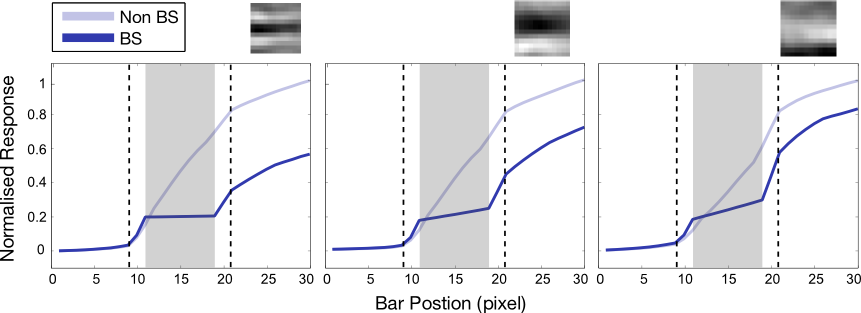}
\end{center}
\caption{{\bf Nonlinear response in BS region at level 1:}
Plots of the normalized absolute value of response are shown against the bar position for three highly active neurons  (indicated by red arrows in the sixth bar blot of figure~\ref{Response_Barshift}) In these plots, dotted rectangular area indicates the extension of BS module whereas, the solid gray rectangular area indicates the extension of blind spot. The receptive fields of these three neurons are shown at the top of the respective plots, which show that these neurons participated in encoding information of a horizontal bar. }
 \label{Response_nonliear}
\end{figure}

Figure(\ref{Response_Barshift}) shows the bar plots of the response of 64 neurons in BS module  at level 1 for six different bar position in both model networks(BS and non-BS). The comparison shows that almost the same set of a small number of neurons responded in both networks. The receptive field of highly responsive neurons in this set possesses a horizontal bar-like structures. We plotted the response of some of these highly responsive neurons against the bar position (varying by one pixel) ~ (see fig. \ref{Response_nonliear}) which show that these neurons  exhibit non-linearly elevated response when bar crosses the blind spot. This nature of the nonlinear behavior as well as the final elevated response are in line with the experimental results~ \cite{matsumoto2005neural}. The response of neurons in BS network is elevated reasonably close to the responses of those neurons in the non-BS network. The closeness of responses indicates the representation of objects in the BS network is similar to the one in the non-BS network. This is reflected in the corresponding ``perceptual images" ( see figure \ref{Bar_shift}b) reconstructed using the generative process.

It is evident from these results that, in the case of BS Network, as long as the bar end reside inside the blind spot the response of neurons, in the BS module, remained constant and relatively low (fig. \ref{Response_Barshift} and fig.\ref{Response_nonliear}) which results in the perception of a bar of constant length on one side of the blind spot. But as the bar crosses the blind spot their responses significantly elevated and the filling-in completion occurred. The elevation in response depends on the localization and the profile of the receptive field of neurons in BS area. These  could also be understood by observing the relative deviation of the response of each neuron (typically the highly responsive neurons) in both networks (fig.\ref{Response_Barshift}). These results are consistent with the findings of neurophysiological studies with bar stimuli on macaque monkeys. 

\begin{figure}
\begin{center}
\includegraphics[width=5.3in]{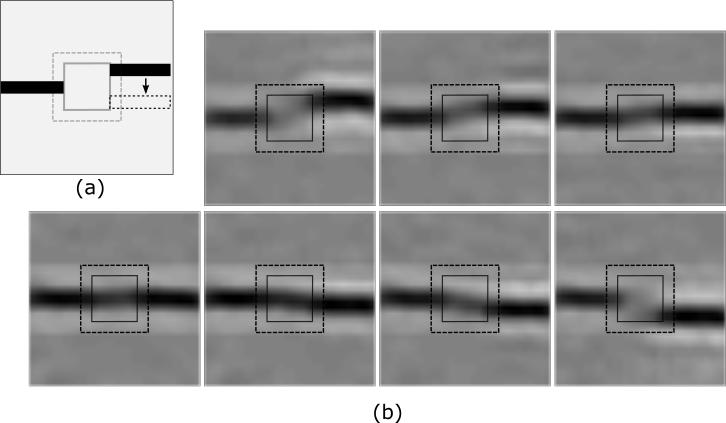}
\end{center}
\caption{{\bf Nonaligned bar investigation.}
a) A typical stimulus is shown, where two non-aligned bar segments are presented at opposite sides of the blind spot. For our study, one bar was fixed (left one) and the other one was shifted vertically by one pixel per instant for the seven positions emulating seven stimuli. b) The generated ``perceptual images'' for those stimuli (as discussed in (a)) corresponding to the recorded response profile of PE neurons at level 1} 
\label{Nonaligned_Bar}
\end{figure}

\paragraph*{ Filling-in for misaligned bars :} Two bar segments were presented on the opposite side of the blind spot. One of those bar segments was kept fixed, whereas another one was shifted along the vertical direction (see figure(\ref{Nonaligned_Bar}a). We recorded the response of PE neurons, in the BS module, at Level 1 and generated the "perceptual images'' corresponding to these misaligned bars. The result presented in the figure(\ref{Nonaligned_Bar}b) shows that the bar appears completed when both segments are aligned, but this filling-in fades away when misalignment increases. This result indicates that the filling-in completion is highly favorable for perfect alignment and have some degree of tolerance against the misalignment. Similar results are reported in earlier psychophysical studies ~\cite{araragi2008anisotropy,araragi2011anisotropies}.

To understand the mechanism of filling-in, we should recall that in HPC, feed-forward connection propagates up the residual error,  corresponding to prediction  made by higher area using generative model of natural images, to correct the current estimate for the  betterment of the next prediction. The optimum-estimate, where prediction closely matches the ``driving sensory'' input as well as ``contextual signal'' from higher area, which produce a minimum prediction error, is then depicted as a percept of the input. But the  blind spot is characterized by the absence of such feed-forward connection that is responsible for propagating up the error. Therefore, the estimate made by higher area, on the basis of surrounding information and the learned natural image statistics,  prevails and this provides the ground for the filling-in completion at the blind-spot.

The estimate at level 2, which operate on a larger spatial scale than the blind spot, is made in a larger context. In such context, an optimum-estimate (or prediction) will be the one that matches the ``driving sensory" input  in the surrounding region of the blind spot, which results in the minimal error signal. Since this estimate is made on the basis of the learned statistical regularity of natural image, the stimuli which have similar statistics in the surrounding region of the blind spot is estimated as continuous regular object across the blind spot, which ultimately results in completion at level 2. In the absence of feed-forward connection in BS region, the corresponding local optimum-estimate at level 1 will, therefore, evolve by matching the ``context driven" feedback signal from level 2. This process at level 1 locally captures all the course of the completion process at level 2. Thus, the properties of the filling-in are highly determined by the matching of statistics of input stimuli around the blind spot and natural statistics learned by the network. The higher degree of matching leads to higher chances of completion.

For example, in the bar shifting experiment while one  end the bar end resides inside the blind spot, the incoming sensory input, which is of a short bar residing on one side of the blind spot, deviates reasonably from learned statistical regularity, in which the bars are usually longer (extended across the blind spot)~\cite{rao1999predictive}. That turns out as a non-completion of the bar and PE neurons at BS module, whose response represent bar, exhibit low response. On the other hand, when bar crosses the blind spot, the  probability of matching with the learned statistical regularity and this abruptly elevates the response of PE neurons which encodes the bar in BS module. This process resembles the AND-gate functionality and reflects the nonlinearity in response profiles shown above (fig ~\ref{Response_nonliear}).

When two aligned bar segments are presented on opposite sides of the blind spot, level 2 estimate (or predict) those bar segments as parts of the single continuous bar extended across the blind spot.  Since there is no feed-foreword corrective information to correct the estimate at level 2 corresponding to blind spot, an estimated long continuous bar extended across the blind spot provides the best match with aligned bar segments as well as with the learned statistical regularity. This process results in the perception of a long continuous bar. On the other hand, two non-aligned such bar is less likely to be a continuous object in the context of the natural image and, hence, their likelihood of completion. The findings of our study suggest that the learned statistical regularity of natural objects along with one way prediction mechanism plays a significant role in filling-in completion.

\section*{Discussion}

The aim of this study was to investigate the computational mechanism related to the filling-in at the blind spot. We speculated that this can be understood in the framework of HPC. We conducted simulation studies on the three level HPC model network to investigate the filling-in completion using bar stimuli. In the first study, we recorded the response of PE neurons at V1 in BS module, while shifting a long bar across the blind spot, and generated the corresponding perceptual counterpart using the generative model. The recorded response shows similar non-linear profile and represents the filled up segment of the bar which is reported by Matsumoto and Komatsu~\cite{matsumoto2005neural}. In the other study, to verify the tolerance of perception of completion of a non-aligned bar, a pair of bar segments were presented on the opposite side of the blind spot, with varying alignment. We found that the filling-in completion, which occurs in the case of perfectly aligned bar segments, also occurs with small degree of misalignment, but as the misalignment increases, the completion doesn't happen. These results are in good agreement with the physiological and psychophysical results reported earlier~\cite{matsumoto2005neural,araragi2011anisotropies}.
 
Previous studies have suggested hierarchical predictive processing of natural images, as a general unified coding principle of visual system~\cite{rao1999predictive,jehee2009predictive,jehee2006learning,hohwy2008predictive,ballard2012dynamic}. This study suggests that the same coding principle could also account for filling-in at the blind spot. Where, for an input stimulus around the blind spot, higher areas (V2) generates unified estimate (which also incorporate the estimate corresponding to blind spot) of the input stimuli on the basis of the learned regularities from natural image statistics. The completion phenomena is the BS region are then simply the outcome of remaining such contextually made, natural image statistics biased estimate uncorrected due to the absence of error carrying feed-forward connection in BS region. The BS region at V1 takes part in this process by not propagating up any error signal and achieving local optimum-estimate by matching contextual and statistically influenced top-down prediction. The achieved multilevel optimum-estimate, comprising a continuous visual object across the blind spot  result in a percept of completion. The nonlinearity observed in responses and, hence, the properties of filling-in, result from the degree of similarity between statistics of stimuli around the blind spot and the natural image statistics. This study provides another support to the suggestions of predictive coding  as a general computational principle of visual cortex.

Studies~\cite{neumann2001visual,grossberg1999neural}, have suggested the role of cortico-cortical (V2-V1) interaction in the filling-in of illusory contours and the surfaces. Neumann~\cite{neumann1999recurrent} suggested that the filling-in of illusory contour could be the outcome of modulation mechanism of feedback signal from V2, which enhance the favorable response profile of feature detecting neurons, mainly in the superficial layers, at V1, in the context of larger contour coded at V2. This model, therefore, has its limitation in explaining the completion across the blind spot where activity is mainly found in the deep layer of the V1. In another recent study \cite{hosoya2012multinomial},authors tried to explain the non-linear behavior of neurons in filling-in in terms of interaction of top-down and bottom-up signal in a Bayesian framework, where the feed-forward signal carrying the ``prediction match'' plays a crucial role. However, in this study, we have demonstrated similar response profiles along with the other properties of filling-in under a simple, unifying framework of hierarchical predictive processing, where feed-forward signal carries the ``prediction error''in spite of ``prediction match" which determines the activities of PE neurons. These PE neurons, in this framework, hypothetically, resides in the deep layer of the cortex~\cite{rao2000predictive} which is consistent with the physiological findings. 

This study does not reject any possible role of intra-cortical interaction in V1 in filling-in completion. There could be some other (or more than one) prediction-correction pathway within V1 which can contribute to filling-in on the basis of contextual information surrounding the blind spot. In this study, we use a simple linear hierarchical predictive coding model network, which deals with static images. The inclusion of a Kalman filter, which can implement spatiotemporal prediction-correction mechanism,  could account for the motion related filling-in properties (such as \cite{murakami1995motion}). Moreover, in our study, we have focused only on filling-in completion of a bar, which is primarily based on the learned statistics on  contrast information (edge, boundary, etc.) found in natural scenes. The surface filling-in at the blind spot could also be understood under the similar mechanism, given a proper surface representation (learning) in the hierarchical probabilistic framework, which is the present challenge of visual science. The future research along these directions could  provide the basis for HPC to explain the complete phenomena of filling-in.

Finally, our study suggests that the filling-in could be a manifestation of a hierarchical predictive process which is, recently, argued to be a general and unifying coding principle of visual cortex. 

\section*{Acknowledgements:}
We thank Rajesh P N Rao, Director of the NSF Center for Sensorimotor Neural Engineering and Professor of Computer Science and Engineering at the University of Washington in Seattle, for valuable discussions and suggestions regarding this work. This work was supported by research grants from the Department of Atomic Energy (DAE), Govt of India.

\bibliographystyle{unsrt}
\bibliography{Full_Paper_corrected_3}

\end{document}